\documentclass[a4paper]{article}
\usepackage{multirow}
\usepackage{INTERSPEECH2019}

\usepackage{amssymb}
\usepackage{amsmath}
\usepackage{xfrac}

\usepackage[dvipsnames]{xcolor}
\newcommand{\be}{\begin{equation}}
\newcommand{\ee}{\end{equation}}
\newcommand{\bea}{\begin{equation} \begin{aligned}}
\newcommand{\eea}{\end{aligned} \end{equation}}

\newcommand{\bln}{\begin{align}}
\newcommand{\eln}{\end{align}}
\newcommand{\bst}{\begin{split}}
\newcommand{\est}{\end{split}}
\newcommand{\bi}{\begin{itemize}}
\newcommand{\ei}{\end{itemize}}
\newcommand{\ben}{\begin{enumerate}}
\newcommand{\een}{\end{enumerate}}

\newcommand{\bprop}{\begin{proposition}}
\newcommand{\eprop}{\end{proposition}}

\definecolor{darkgreen}{rgb}{0,0.6,0}

\title{SpecAugment: A Simple Data Augmentation Method\\
for Automatic Speech Recognition}
\name{Daniel S. Park\sthanks{~~Work done as a member of the
Google AI Residency Program.},~
  William Chan,
  Yu Zhang,
  Chung-Cheng Chiu,\\
  Barret Zoph,
  Ekin D. Cubuk,
  Quoc V. Le
  }
\address{
Google Brain}
\email{\{danielspark, williamchan, ngyuzh, chungchengc, barretzoph, cubuk, qvl\}@google.com}

\begin{document}

\maketitle
\begin{abstract}
We present SpecAugment, a simple data augmentation method for speech recognition. SpecAugment is applied directly to the feature inputs of a neural network (i.e., filter bank coefficients). The augmentation policy consists of warping the features, masking blocks of frequency channels, and masking blocks of time steps. We apply SpecAugment on Listen, Attend and Spell networks for end-to-end speech recognition tasks. We achieve state-of-the-art performance on the LibriSpeech 960h and Swichboard 300h tasks, outperforming all prior work. On LibriSpeech, we achieve 6.8\% WER on test-other without the use of a language model, and 5.8\% WER with shallow fusion with a language model. This compares to the previous state-of-the-art hybrid system of 7.5\% WER. For Switchboard, we achieve 7.2\%/14.6\% on the Switchboard/CallHome portion of the Hub5'00 test set without the use of a language model, and 6.8\%/14.1\% with shallow fusion, which compares to the previous state-of-the-art hybrid system at 8.3\%/17.3\% WER.
\end{abstract}

\section{Introduction}

Deep Learning has been applied successfully to Automatic Speech Recognition (ASR)~\cite{hinton2012deep}, where the main focus of research has been  designing better  network architectures, for example, DNNs \cite{dahl-ieeetasl-2012}, CNNs \cite{sainath-icassp-2013}, RNNs \cite{graves-icassp-2013} and end-to-end models \cite{graves-icml-2014,Chan2016ListenAA,bahdanau-icassp-2016}. However, these  models tend to overfit easily and require large amounts of training data \cite{chiu-icassp-2018}.

Data augmentation has been proposed as a method to generate additional training data for ASR. For example, in \cite{kanda-asru-2013,ragni-interspeech-2014}, artificial data was augmented for low resource speech recognition tasks. Vocal Tract Length Normalization has been adapted for data augmentation in \cite{jaitly-2013-icml}. Noisy audio has been synthesised via superimposing clean audio with a noisy audio signal in \cite{hannun-arxiv-2014}. Speed perturbation has been applied on raw audio for LVSCR tasks in \cite{ko-interspeech-2015}. The use of an acoustic room simulator has been explored in \cite{kim-interspeech-2017}. Data augmentation for keyword spotting has been studied in \cite{prabhavalkar-2015-icassp, raju-arxiv-2018}. Feature drop-outs have been employed for training multi-stream ASR systems \cite{mallidi-icassp-2016}. More generally, learned augmentation techniques have explored different sequences of augmentation transformations that have achieved state-of-the-art performance in the image domain \cite{Cubuk2018AutoAugmentLA}.

Inspired by the recent success of augmentation in the speech and vision domains, we propose SpecAugment, an augmentation method that operates on the log mel spectrogram of the input audio, rather than the raw audio itself. This method is simple and computationally cheap to apply, as it directly acts on the log mel spectrogram as if it were an image, and does not require any additional data. We are thus able to apply SpecAugment online during training. SpecAugment consists of three kinds of deformations of the log mel spectrogram. The first is time warping, a deformation of the time-series in the time direction. The other two augmentations, inspired by ``Cutout", proposed in computer vision \cite{devries-arxiv-2017}, are time and frequency masking, where we mask a block of consecutive time steps or mel frequency channels.

This approach while rudimentary, is remarkably effective and allows us to train end-to-end ASR networks, called Listen Attend and Spell (LAS) \cite{Chan2016ListenAA}, to surpass more complicated hybrid systems, and achieve state-of-the-art results even without the use of Language Models (LMs). 
On LibriSpeech \cite{Panayotov2015LibriSpeechAA}, we achieve 2.8\% Word Error Rate (WER) on the test-clean set and 6.8\% WER on the test-other set, without the use of an LM. Upon shallow fusion \cite{gulcehre-2015-arxiv} with an LM trained on the LibriSpeech LM corpus, we are able to better our performance (2.5\% WER on test-clean and 5.8\% WER on test-other), improving the current state of the art on test-other by 22\% relatively. On Switchboard 300h (LDC97S62) \cite{switchboard}, we obtain 7.2\% WER on the Switchboard portion of the Hub5'00 (LDC2002S09, LDC2003T02) test set, and 14.6\% on the CallHome portion, without using an LM. Upon shallow fusion with an LM trained on the combined transcript of the Switchboard and Fisher (LDC200\{4,5\}T19) \cite{fisher} corpora, we obtain 6.8\%/14.1\% WER on the Switchboard/Callhome portion.

\section{Augmentation Policy}

\begin{figure}[t]
  \centering
  \begin{tabular}{c}
  \includegraphics[height=1.0cm]{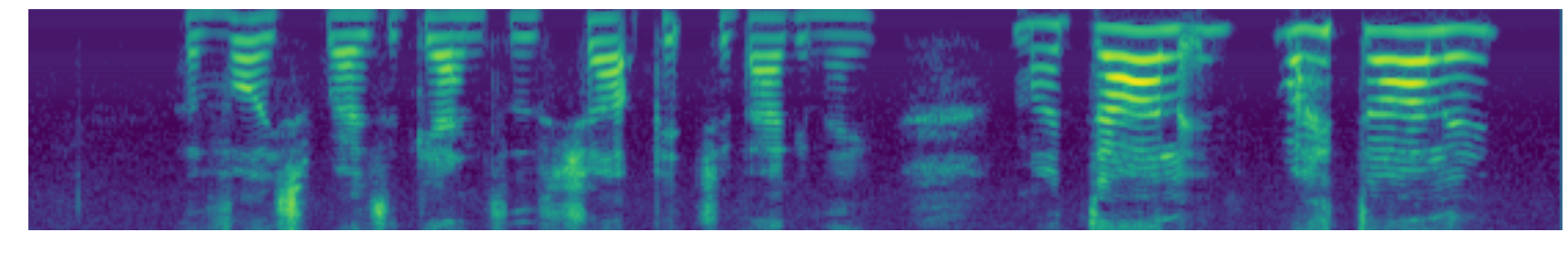} \\
  \includegraphics[height=1.0cm]{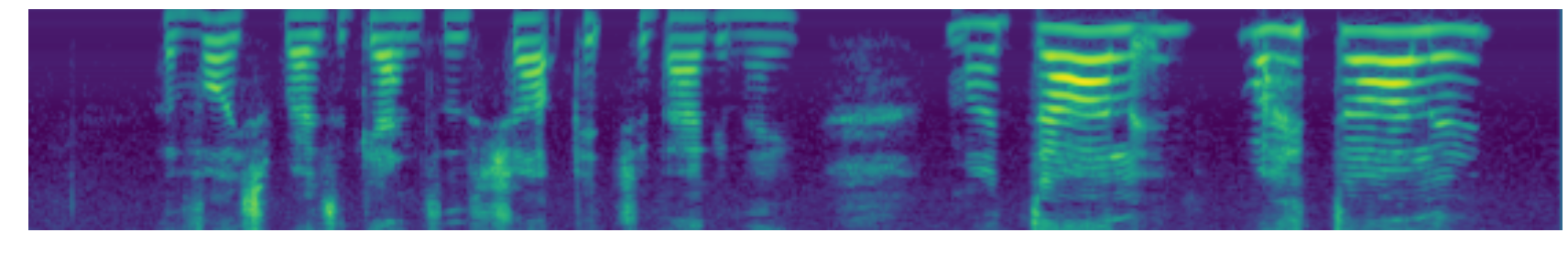} \\
  \includegraphics[height=1.0cm]{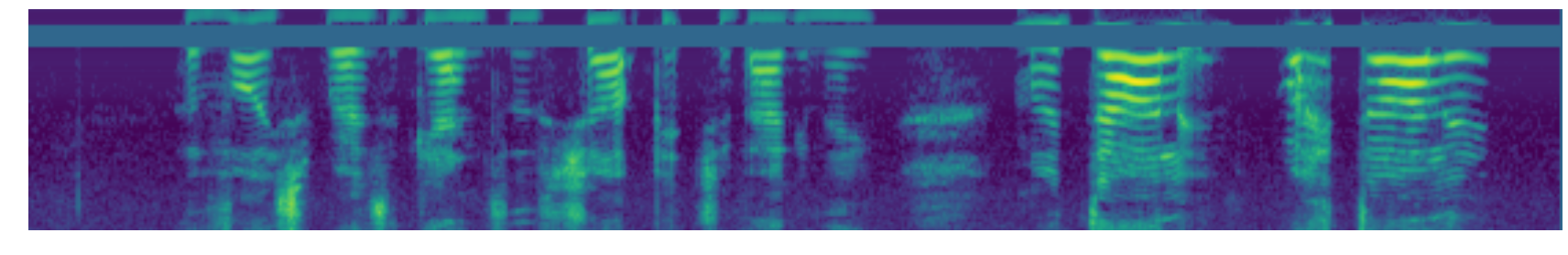} \\
  \includegraphics[height=1.0cm]{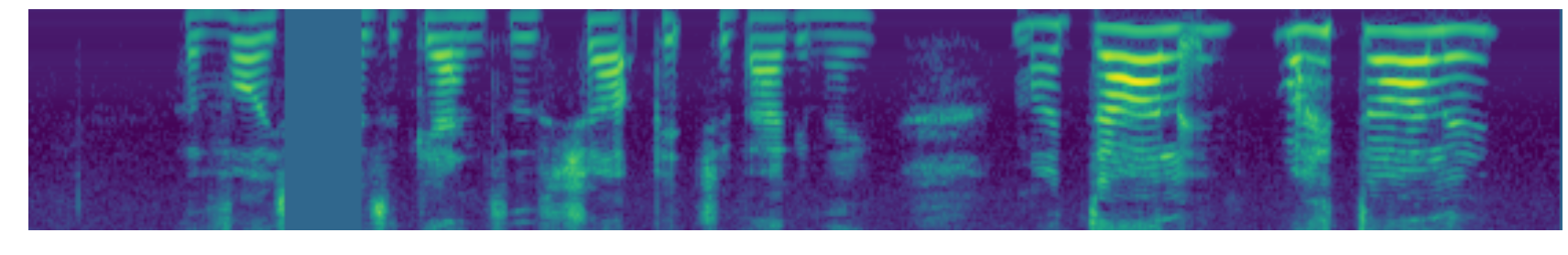}
  \end{tabular}
  \vskip -0.1in
  \caption{Augmentations applied to the base input, given at the top. From top to bottom, the figures depict the log mel spectrogram of the base input with no augmentation, time warp, frequency masking and time masking applied.}
  \label{fig:augs}
\end{figure}

We aim to construct an augmentation policy that acts on the log mel spectrogram directly, which helps the network learn useful features. Motivated by the goal that these features should be robust to deformations in the time direction, partial loss of frequency information and partial loss of small segments of speech, we have chosen the following deformations to make up a policy:
\begin{enumerate}
\item Time warping is applied via the function {\tt sparse\_image\_warp} of {\tt tensorflow}. Given a log mel spectrogram with $\tau$ time steps, we view it as an image where the time axis is horizontal and the frequency axis is vertical. A random point along the horizontal line passing through the center of the image within the time steps $(W, \tau-W)$ is to be warped either to the left or right by a distance $w$ chosen from a uniform distribution from 0 to the time warp parameter $W$ along that line.  We fix six anchor points on the boundary---the four corners and the mid-points of the vertical edges.
\item Frequency masking is applied so that $f$ consecutive mel frequency channels $[f_0, f_0 + f)$ are masked, where $f$ is first chosen from a uniform distribution from 0 to the frequency mask parameter $F$, and $f_0$ is chosen from $[0, \nu -f)$. $\nu$ is the number of mel frequency channels.
\item Time masking is applied so that $t$ consecutive time steps $[t_0, t_0 + t)$ are masked, where $t$ is first chosen from a uniform distribution from 0 to the time mask parameter $T$, and $t_0$ is chosen from $[0, \tau-t)$.
 \begin{itemize}
 \item We introduce an upper bound on the time mask so that a time mask cannot be wider than $p$ times the number of time steps.
 \end{itemize}
\end{enumerate}
Figure \ref{fig:augs} shows examples of the individual augmentations applied to a single input. The log mel spectrograms are normalized to have zero mean value, and thus setting the masked value to zero is equivalent to setting it to the mean value.

\begin{figure}[t]
  \centering
  \begin{tabular}{c}
  \includegraphics[height=1.0cm]{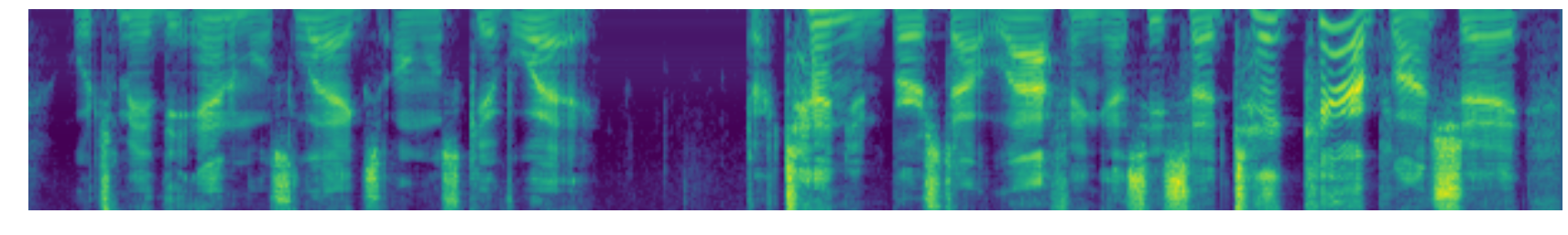} \\
  \includegraphics[height=1.0cm]{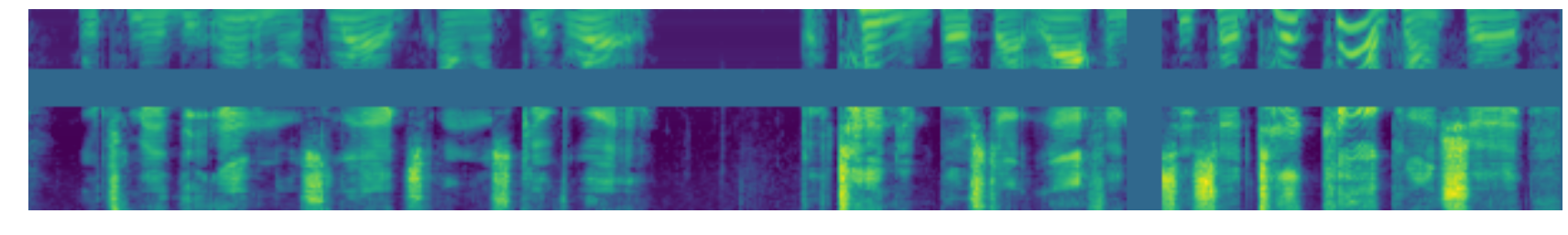} \\
  \includegraphics[height=1.0cm]{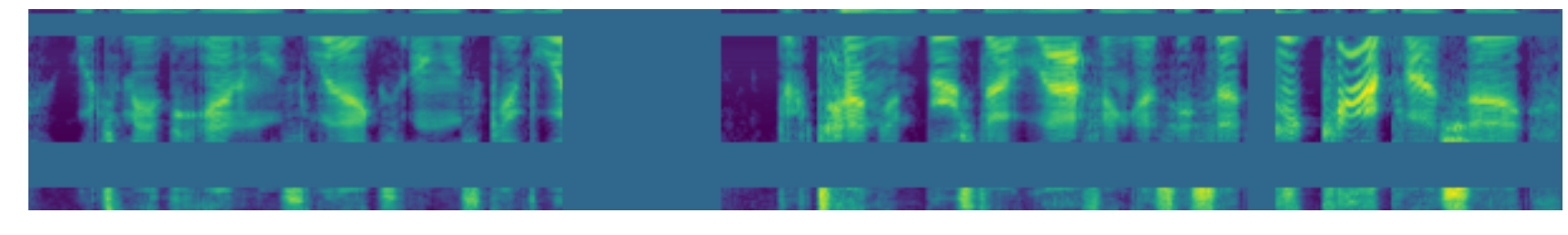}
  \end{tabular}
  \vskip -0.05in
  \caption{Augmentation policies applied to the base input. From top to bottom, the figures depict the log mel spectrogram of the base input with policies None, LB and LD applied.}
  \label{fig:policies}
  \vskip -0.15in
\end{figure}

We can consider policies where multiple frequency and time masks are applied. The multiple masks may overlap. In this work, we mainly consider a series of hand-crafted policies, LibriSpeech basic (LB), LibriSpeech double (LD), Switchboard mild (SM) and Switchboard strong (SS) whose parameters are summarized in Table \ref{t:policies}. In Figure \ref{fig:policies}, we show an example of a log mel spectrogram augmented with policies LB and LD.

\begin{table}[h]
  \caption{Augmentation parameters for policies. $m_F$ and $m_T$ denote the number of frequency and time masks applied.}
  \label{t:policies}
  \centering
  \footnotesize
  \begin{tabular}{ccccccc}
    \toprule
    Policy & $W$ & $F$ & $m_F$ & $T$ & $p$ & $m_T$ \\
    \midrule
    None & 0 & 0 & - & 0 & - & - \\
    LB & 80 & 27 & 1 & 100 & 1.0 & 1\\
    LD & 80 & 27 & 2 & 100 & 1.0 & 2 \\
    SM & 40 & 15 & 2 & 70 & 0.2 & 2\\
    SS & 40 & 27 & 2 & 70 & 0.2 & 2\\
    \bottomrule
  \end{tabular}
  \vskip -0.1in
\end{table}

\section{Model}

We use Listen, Attend and Spell (LAS) networks \cite{Chan2016ListenAA} for our ASR tasks. These models, being end-to-end, are simple to train and have the added benefit of having well-documented benchmarks \cite{zeyer-interspeech-2018, irie-arxiv-2019} that we are able to build upon to get our results. In this section, we review LAS networks and introduce some notation to parameterize them. We also introduce the learning rate schedules we use to train the networks, as they turn out to be an important factor in determining performance. We end with reviewing shallow fusion \cite{gulcehre-2015-arxiv}, which we have used to incorporate language models for further gains in performance.

\subsection{LAS Network Architectures}

We use Listen, Attend and Spell (LAS) networks \cite{Chan2016ListenAA} for end-to-end ASR studied in \cite{irie-arxiv-2019}, for which we use the notation LAS-$d$-$w$. The input log mel spectrogram is passed in to a 2-layer Convolutional Neural Network (CNN) with max-pooling and stride of $2$. The output of the CNN is passes through an encoder consisting of $d$ stacked bi-directional LSTMs with cell size $w$ to yield a series of attention vectors. The attention vectors are fed into a 2-layer RNN decoder of cell dimension $w$, which yields the tokens for the transcript. The text is tokenized using a Word Piece Model (WPM) \cite{schuster-icassp-2012} of vocabulary size 16k for LibriSpeech and 1k for Switchboard. The WPM for LibriSpeech 960h is constructed using the training set transcripts. For the Switchboard 300h task, transcripts from the training set are combined with those of the Fisher corpus to construct the WPM. The final transcripts are obtained by a beam search with beam size 8. For comparison with \cite{irie-arxiv-2019}, we note that their ``large model" in our notation is LAS-4-1024.

\subsection{Learning Rate Schedules}

The learning rate schedule turns out to be an important factor in determining the performance of ASR networks, especially so when augmentation is present. Here, we introduce training schedules that serve two purposes. First, we use these schedules to verify that a longer schedule improves the final performance of the network, even more so with augmentation (Table \ref{t:LibriSpeech}). Second, based on this, we introduce very long schedules that are used to maximize the performance of the networks.

We use a learning rate schedule in which we ramp-up, hold, then exponentially decay the learning rate until it reaches $\sfrac{1}{100}$ of its maximum value. The learning rate is kept constant beyond this point. This schedule is parameterized by three time stamps $(s_r, s_i, s_f)$ -- the step $s_r$ where the ramp-up (from zero learning rate) is complete, the step $s_i$ where exponential decay starts, and the step $s_f$ where the exponential decay stops.

There are two more factors that introduce time scales in our experiment. First, we turn on a variational weight noise \cite{graves-nips-2011} of standard deviation 0.075 at step $s_\text{noise}$ and keep it constant throughout training. Weight noise is introduced in the step interval $(s_r, s_i)$, i.e., during the high plateau of the learning rate.

Second, we introduce uniform label smoothing \cite{lbsm} with uncertainty 0.1, i.e., the correct class label is assigned the confidence 0.9, while the confidence of the other labels are increased accordingly. As is commented on again later on, label smoothing can destabilize training for smaller learning rates, and we sometimes choose to turn it on only at the beginning of training and off when the learning rate starts to decay.

The two basic schedules we use, are given as the following:
\begin{enumerate}
\item B(asic): $(s_r, s_\text{noise}, s_i, s_f)$ = (0.5k, 10k, 20k, 80k)
\item D(ouble): $(s_r, s_\text{noise}, s_i, s_f)$ = (1k, 20k, 40k, 160k)
\end{enumerate}
As discussed further in section \ref{s:discussion}, we can improve the performance of the trained network by using a longer schedule. We thus introduce the following schedule:
\begin{enumerate}
  \setcounter{enumi}{2}
  \item L(ong): $(s_r, s_\text{noise}, s_i, s_f)$ = (1k, 20k, 140k, 320k)
\end{enumerate}
which we use to train the largest model to improve performance. When using schedule L, label smoothing with uncertainty $0.1$ is introduced for time steps $< s_i = \text{140k}$ for LibriSpeech 960h, and is subsequently turned off. For Switchboard 300h, label smoothing is turned on throughout training.

\subsection{Shallow Fusion with Language Models}

While we are able to get state-of-the-art results with augmentation, we can get further improvements by using a language model. We thus incorporate an RNN language model by shallow fusion for both tasks. In shallow fusion, the ``next token" $\mathbf{y}^*$ in the decoding process is determined by
\be
\mathbf{y}^* = \underset{\mathbf{y}}{\text{argmax}} \left( \log P(\mathbf{y} | \mathbf{x}) + \lambda \log P_{LM} (\mathbf{y})\right) \,,
\ee
i.e., by jointly scoring the token using the base ASR model and the language model. We also use a coverage penalty $c$ \cite{chorowski-2017-interspeech}.

For LibriSpeech, we use a two-layer RNN with embedding dimension 1024 used in \cite{irie-arxiv-2019} for the LM, which is trained on the LibriSpeech LM corpus. We use identical fusion parameters ($\lambda =0.35$ and $c=0.05$) used in \cite{irie-arxiv-2019} throughout.

For Switchboard, we use a two-layer RNN with embedding dimension 256, which is trained on the combined transcripts of the Fisher and Switchboard datasets. We find the fusion parameters via grid search by measuring performance on RT-03 (LDC2007S10). We discuss the fusion parameters used in individual experiments in section \ref{ss:swbd}.

\section{Experiments}

In this section, we describe our experiments on LibriSpeech and Switchboard with SpecAugment. We report state-of-the-art results that out-perform heavily engineered hybrid systems.

\subsection{LibriSpeech 960h} \label{ss:librispeech}

For LibriSpeech, we use the  same setup as \cite{irie-arxiv-2019}, where we use 80-dimensional filter banks with delta and delta-delta acceleration, and a 16k word piece model \cite{schuster-icassp-2012}.

The three networks LAS-4-1024, LAS-6-1024 and LAS-6-1280 are trained on LibriSpeech 960h with a combination of augmentation policies (None, LB, LD) and training schedules (B/D). Label smoothing was not applied in these experiments. The experiments were run with peak learning rate of $0.001$ and batch size of 512, on 32 Google Cloud TPU chips for 7 days. Other than the augmentation policies and learning rate schedules, all other hyperparameters were fixed, and no additional tuning was applied. We report test set numbers validated by the dev-other set in Table \ref{t:LibriSpeech}. We see that augmentation consistently improves the performance of the trained network, and that the benefit of a larger network and a longer learning rate schedule is more apparent with harsher augmentation.

\begin{table}[h!]
  \caption{LibriSpeech test WER (\%) evaluated for varying networks, schedules and policies. First row from \cite{irie-arxiv-2019}.}
  \footnotesize
  \label{t:LibriSpeech}
  \centering
  \resizebox{\columnwidth}{!}{%
  \begin{tabular}{cccrrrr}
    \toprule
    \multirow{2}{*}{\bfseries Network} & \multirow{2}{*}{\bfseries Sch} & \multirow{2}{*}{\bfseries Pol}
    & \multicolumn{2}{c}{\bfseries No LM}  & \multicolumn{2}{c}{\bfseries With LM} \\
    \cmidrule(r){4-5} \cmidrule(r){6-7}
    & & & {\bfseries clean} & {\bfseries other} & {\bfseries clean} & {\bfseries other} \\
    \midrule
    {LAS-4-1024} \cite{irie-arxiv-2019}  & B & - & 4.7 & 13.4 & 3.6 & 10.3 \\
    \midrule
    \multirow{5}{*}{LAS-4-1024} 
    & B & LB & 3.7 & 10.0 & 3.4 & 8.3 \\
    & B & LD & 3.6 & 9.2 & 2.8 & 7.5 \\
    \cmidrule(r){2-7}
    & D & - &  4.4 &  13.3 & 3.5 & 10.4 \\
    & D & LB & 3.4 & 9.2 & 2.7 & 7.3 \\
    & D & LD & 3.4 & 8.3 & 2.8 & 6.8  \\
    \midrule
    \multirow{3}{*}{LAS-6-1024} & D & - & 4.5 & 13.1 & 3.6 & 10.3 \\
    & D & LB & 3.4 & 8.6 & 2.6 & 6.7  \\
    & D & LD & 3.2 & 8.0 & 2.6 & 6.5 \\
    \midrule
    \multirow{3}{*}{LAS-6-1280} & D & - & 4.3 & 12.9 & 3.5 & 10.5  \\
    & D & LB & 3.4 & 8.7 & 2.8 & 7.1  \\
    & D & LD & 3.2 & 7.7 & 2.7 & 6.5  \\
    \bottomrule
  \end{tabular}
  }
\end{table}

We take the largest network, LAS-6-1280, and use schedule L (with training time $\sim$ 24 days) and policy LD to train the network to maximize performance. We turn label smoothing on for time steps $< 140$k as noted before. The test set performance is reported by evaluating the checkpoint with best dev-other performance. State of the art performance is achieved by the LAS-6-1280 model, even without a language model. We can incorporate an LM using shallow fusion to further improve performance. The results are presented in Table \ref{t:libriSOTA}.

\begin{table}[h!]
  \caption{LibriSpeech 960h WERs (\%).}
  \label{t:libriSOTA}
  \centering
  \small
  \resizebox{\columnwidth}{!}{%
  \begin{tabular}{lcccc}
    \toprule
    \bfseries Method & \multicolumn{2}{c}{\bfseries No LM} & \multicolumn{2}{c}{\bfseries With LM} \\
    \cmidrule(r){2-3} \cmidrule(r){4-5}
     & \bfseries clean & \bfseries other & \bfseries clean & \bfseries other \\
    \midrule
    \bfseries HMM \\
    \quad Panayotov et al., (2015) \cite{Panayotov2015LibriSpeechAA} & & & 5.51 & 13.97 \\
    \quad Povey et al., (2016) \cite{povey-interspeech-2016} & & & 4.28 \\
    \quad Han et al., (2017) \cite{han-arxiv-2017} & & & 3.51 &  8.58 \\
    \quad Yang et al. (2018) \cite{yang-arxiv-2018} & & & 2.97 & 7.50 \\
    \midrule
    \bfseries CTC/ASG \\
    \quad Collobert et al., (2016) \cite{collobert-arxiv-2016} & 7.2 \\
    \quad Liptchinsky et al., (2017) \cite{liptchinsky-arxiv-2017} & 6.7 & 20.8 & 4.8 & 14.5 \\
    \quad Zhou et al., (2018) \cite{zhou-icassp-2018} & & & 5.42 & 14.70 \\
    \quad Zeghidour et al., (2018) \cite{zeghidour-arxiv-2018} & & & 3.44 & 11.24 \\
    \quad Li et al., (2019) \cite{li-arxiv-2019} & 3.86 & 11.95 & 2.95 & 8.79 \\
    \midrule
    \bfseries LAS \\
    \quad Zeyer et al., (2018) \cite{zeyer-interspeech-2018} & 4.87 & 15.39 & 3.82 & 12.76 \\
    \quad Zeyer et al., (2018) \cite{zeyer-nips-2018} & 4.70 & 15.20 \\
    \quad Irie et al., (2019) \cite{irie-arxiv-2019} & 4.7 & 13.4 & 3.6 & 10.3 \\
    \quad Sabour et al., (2019) \cite{sabour-iclr-2019} & 4.5 & 13.3 \\
    \midrule
    \bfseries Our Work \\
    \quad LAS & 4.1 & 12.5 & 3.2 & 9.8 \\
    \quad LAS + SpecAugment & \bfseries 2.8 & \bfseries 6.8 & \bfseries 2.5 & \bfseries 5.8 \\
    \bottomrule
  \end{tabular}
  }
  \vskip -0.1in
\end{table}

\subsection{Switchboard 300h} \label{ss:swbd}

For Switchboard 300h, we use the Kaldi \cite{povey-asru-2011} ``s5c'' recipe to process our data, but we adapt the recipe to use 80-dimensional filter banks with delta and delta-delta acceleration. We use a 1k WPM \cite{schuster-icassp-2012} to tokenize the output, constructed using the combined vocabulary of the Switchboard and Fisher transcripts.

We train LAS-4-1024 with policies (None, SM, SS) and schedule B. As before, we set the peak learning rate to $0.001$ and total batch size to 512, and train using 32 Google Cloud TPU chips. Here the experiments are run with and without label smoothing. Not having a canonical development set, we choose to evaluate the checkpoint at the end point of the training schedule, which we choose to be 100k steps for schedule B. We note that the training curve relaxes after the decay schedule is completed (step $s_f$), and the performance of the network does not vary much. The performance of various augmentation policies with and without label smoothing for Switchboard 300h is shown in Table \ref{t:SWBD}. We see that label smoothing and augmentation have an additive effect for this corpus.

\begin{table}[h!]
  \vskip -0.1in
  \caption{Switchboard 300h WER (\%) evaluated for LAS-4-1024 trained with schedule B with varying augmentation and Label Smoothing (LS) policies. No LMs have been used.}
  \footnotesize
  \label{t:SWBD}
  \centering
  \begin{tabular}{ccrr}
    \toprule
    {\bfseries Policy} & {\bfseries LS} & {\bfseries SWBD} & {\bfseries CH} \\
    \midrule
    \multirow{2}{*}{-} & $\times$ & 12.1 & 22.6 \\
    & $\circ$ & 11.2 & 21.6 \\
    \midrule
    \multirow{2}{*}{SM}& $\times$ & 9.5 & 18.8 \\
    & $\circ$ & 8.5 & 16.1 \\
    \midrule
    \multirow{2}{*}{SS} & $\times$ & 9.7 & 18.2 \\
    & $\circ$ & 8.6 & 16.3 \\
    \bottomrule
  \end{tabular}
\end{table}

As with LibriSpeech 960h, we train LAS-6-1280 on the Switchboard 300h training set with schedule L (training time $\sim$ 24 days) to get state of the art performance. In this case, we find that turning label smoothing on throughout training benefits the final performance. We report the performance at the end of training time at 340k steps. We present our results in the context of other work in Table \ref{t:swbdSOTA}.  We also apply shallow fusion with an LM trained on Fisher-Switchboard, whose fusion parameters are obtained by evaluating performance on the RT-03 corpus. Unlike the case for LibriSpeech, the fusion parameters do not transfer well between networks trained differently---the three entries in Table \ref{t:swbdSOTA} were obtained by using fusion parameters $(\lambda, c) = $ (0.3, 0.05), (0.2, 0.0125) and (0.1, 0.025) respectively.

\begin{table}[h!]
  \caption{Switchboard 300h WERs (\%).}
  \label{t:swbdSOTA}
  \centering
  \small
  \resizebox{\columnwidth}{!}{%
  \begin{tabular}{lcccc}
    \toprule
    \bfseries Method & \multicolumn{2}{c}{\bfseries No LM} & \multicolumn{2}{c}{\bfseries With LM} \\
    \cmidrule(r){2-3} \cmidrule(r){4-5}
    & \bfseries SWBD & \bfseries CH & \bfseries SWBD & \bfseries CH \\
    \midrule
    \bfseries HMM \\
    \quad Vesel{\'y} et al., (2013) \cite{vesely-interspeech-2013} & & & 12.9 & 24.5 \\
    \quad Povey et al., (2016) \cite{povey-interspeech-2016} & & & 9.6 & 19.3 \\
    \quad Hadian et al., (2018) \cite {hadian-interspeech-2018} & & & 9.3 & 18.9 \\
    \quad Zeyer et al., (2018) \cite{zeyer-interspeech-2018} & & & 8.3 & 17.3 \\
    \midrule
    \bfseries CTC \\
    \quad Zweig et al., (2017) \cite{zweig-icassp-2017} & 24.7 & 37.1 & 14.0 & 25.3 \\
    \quad Audhkhasi et al., (2018) \cite{audhkhasi-interspeech-2019} & 20.8 & 30.4 \\
    \quad Audhkhasi et al., (2018) \cite{audhkhasi-icassp-2018} & 14.6 & 23.6 \\
    \midrule
    \bfseries LAS \\
    \quad Lu et al., (2016) \cite{lu-icassp-2016} & 26.8 & 48.2 & 25.8 & 46.0 \\
    \quad Toshniwal et al., (2017) \cite{toshniwal-interspeech-2017} & 23.1 & 40.8 \\
    \quad Zeyer et al., (2018) \cite{zeyer-interspeech-2018} & 13.1 & 26.1 & 11.8 & 25.7 \\
    \quad Weng et al., (2018) \cite{weng-interspeech-2018} & 12.2 & 23.3 \\
    \quad Zeyer et al., (2018) \cite{zeyer-nips-2018} & 11.9 & 23.7 & 11.0 & 23.1 \\
    \midrule
    \bfseries Our Work \\
    \quad LAS & 11.2 & 21.6 & 10.9 & 19.4 \\
    \quad LAS + SpecAugment (SM) & \textbf{7.2} & 14.6 & \textbf{6.8} & 14.1 \\
    \quad LAS + SpecAugment (SS) & 7.3 & \textbf{14.4} & 7.1 & \textbf{14.0} \\
    \bottomrule
  \end{tabular}
  }
  \vskip -0.1in
\end{table}

\section{Discussion}\label{s:discussion}

\noindent\textbf{Time warping contributes, but is not a major factor in improving performance.} In Table \ref{tab:abalation}, we present three training results for which time warping, time masking and frequency masking have been turned off, respectively. We see that the effect time warping, while small, is still existent. Time warping, being the most expensive as well as the least influential of the augmentations discussed in this work, should be the first augmentation to be dropped given any budgetary limitations.
\smallskip

\begin{table}[th]
  \vskip -0.1in
  \caption{Test set WER (\%) evaluated without LM for network LAS-4-1024 trained with schedule B.}
  \label{tab:abalation}
  \centering
  \footnotesize
  \begin{tabular}{ccccccrr}
    \toprule
    $W$ & $F$ & $m_F$ & $T$ & $p$ & $m_T$ & test-other & test \\
    \midrule
    80 & 27 & 1 & 100 & 1.0 & 1 & 10.0 & 3.7 \\
    \midrule
    0  & 27 & 1 & 100 & 1.0 & 1 & 10.1 & 3.8 \\
    80 &  0 & - & 100 & 1.0 & 1 & 11.0 & 4.0 \\
    80 & 27 & 1 & 0 & - & - & 10.9 & 4.1 \\
    \bottomrule
  \end{tabular}
  \vskip -0.1in
\end{table}

\noindent\textbf{Label smoothing introduces instability to training.} We have noticed that the proportion of unstable training runs increases for LibriSpeech when label smoothing is applied with augmentation. This becomes more conspicuous while learning rate is being decayed, thus our introduction of a label smoothing schedule for training LibriSpeech, where labels are only smoothed in the initial phases of the learning rate schedule.
\smallskip

\begin{figure}[h]
  \centering
  \begin{tabular}{cc}
  \includegraphics[height=3.3cm]{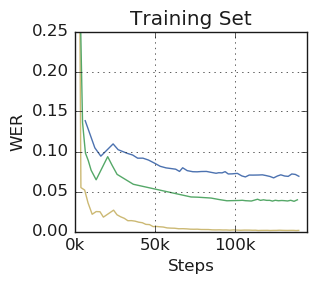} &
  \includegraphics[height=3.3cm]{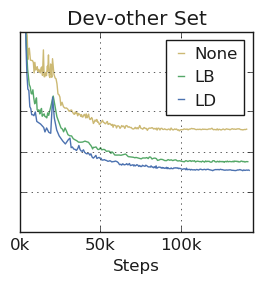}
  \end{tabular}
  \vskip - 0.05in
  \caption{LAS-6-1280 on LibriSpeech with schedule D.}
  \label{f:training}
  \vskip -0.1in
\end{figure}

\noindent\textbf{Augmentation converts an over-fitting problem into an under-fitting problem.} As can be observed from the training curves of the networks in Figure \ref{f:training}, the networks during training not only under-fit the loss and WER on the augmented training set, but also on the training set itself when trained on augmented data. This is in stark contrast to the usual situation where networks tend to over-fit to the training data. This is the major benefit of training with augmentation, as explained below.
\smallskip

\noindent\textbf{Common methods of addressing under-fitting yield improvements.} We were able to make significant gains in performance by standard approaches to alleviate under-fitting---making larger networks and training longer. The current reported performance was obtained by the recursive process of applying a harsh augmentation policy, and then making wider, deeper networks and training them with longer schedules to address the under-fitting.
\smallskip

\noindent\textbf{Remark on related works.} We note that an augmentation similar to frequency masking has been studied in the context of CNN acoustic models in \cite{kovacs2017increasing}. There, blocks of adjacent frequencies are pre-grouped into bins, which are randomly zeroed-out per minibatch. On the other hand, both the size and position of the frequency masks in SpecAugment are chosen stochastically, and differ for every input in the minibatch. More ideas for structurally omitting frequency data of spectrograms have been discussed in \cite{toth2018perceptually}.

\section{Conclusions}

SpecAugment greatly improves the performance of ASR networks. We are able to obtain state-of-the-art results on the LibriSpeech 960h and Switchboard 300h tasks on end-to-end LAS networks by augmenting the training set using simple hand-crafted policies, surpassing the performance of hybrid systems even without the aid of a language model. SpecAugment converts ASR from an over-fitting to an under-fitting problem, and we were able to gain performance by using bigger networks and training longer.
\smallskip

\noindent
\textbf{Acknowledgements:}
We would like to thank Yuan Cao, Ciprian Chelba, Kazuki Irie, Ye Jia, Anjuli Kannan, Patrick Nguyen, Vijay Peddinti, Rohit Prabhavalkar, Yonghui Wu and Shuyuan Zhang for useful discussions. We also thank Gy\"orgy Kov\'acs for introducing us to the works \cite{kovacs2017increasing, toth2018perceptually}.

\bibliographystyle{IEEEtran}

\bibliography{mybib}

\end{document}